# Generalized Kramers-Kronig Receiver for Coherent THz Communications

T. Harter[1,2*], C. Füllner[1*], J. N. Kemal[1], S. Ummethala[1,2], J. L. Steinmann[3], M. Brosi[3], J. L. Hesler[4], E. Bründermann[3], A.-S. Müller[3], W. Freude[1], S. Randel[1**], C. Koos[1,2**]

[1]Institute of Photonics and Quantum Electronics (IPQ), Karlsruhe Institute of Technology (KIT), Germany
[2]Institute of Microstructure Technology (IMT), Karlsruhe Institute of Technology (KIT), Germany
[3]Institute for Beam Physics and Technology (IBPT), Karlsruhe Institute of Technology (KIT), Germany
[4]Virginia Diodes, Inc. (VDI), Charlottesville, United States of America

*These authors contributed equally to this work
**e-mail: sebastian.randel@kit.edu, christian.koos@kit.edu

**High-speed communication systems rely on spectrally efficient modulation formats that encode information both on the amplitude and on the phase of an electromagnetic carrier. Coherent detection of such signals typically uses rather complex receiver schemes, requiring a continuous-wave (c.w.) local oscillator (LO) as a phase reference and a mixer circuit for spectral down-conversion. In optical communications, the so-called Kramers-Kronig (KK) scheme has been demonstrated to greatly simplify the receiver, reducing the hardware to a single photodiode[1–3]. In this approach, an LO tone is transmitted along with the signal, and the amplitude and phase of the complex signal envelope are reconstructed from the photocurrent by digital signal processing. This reconstruction exploits the fact that the real and the imaginary part, or, equivalently, the amplitude and the phase of an analytic signal are connected by a KK-type relation[4–6]. Here, we transfer the KK scheme to high-speed wireless communications at THz carrier frequencies. We use a Schottky-barrier diode (SBD) as a nonlinear element and generalize the theory of KK processing to account for the non-quadratic characteristics of this device. Using 16-state quadrature amplitude modulation (16QAM), we transmit a net data rate of 115 Gbit/s at a carrier frequency of 0.3 THz over a distance of 110 m.**

Future mobile communication networks will crucially rely on wireless backbone networks with high-speed point-to-point links[7,8]. To offer data rates of 100 Gbit/s or more, these links will have to exploit unoccupied spectral resources at carrier frequencies[7,9–14] above 0.1 THz, where frequency windows of low atmospheric attenuation enable transmission over practically relevant distances of a few hundred meters[15,16]. To generate the underlying THz data signals, coherent down-conversion of optical waveforms in high-speed photodiodes has been demonstrated as a particularly promising approach[7,10–14,17–19]. This concept opens a simple path towards quadrature amplitude modulation formats that offer fast transmission at high spectral efficiencies. Coherent reception of these signals, however, largely relies on relatively complex THz circuits, which comprise, e.g., high-speed mixers along with THz LOs. These circuits are costly and often represent the bandwidth bottleneck of the transmission link[10].

In this paper, we demonstrate a greatly simplified coherent receiver scheme for THz data signals that relies on a simple envelope detector and subsequent digital signal processing[20] (DSP). The scheme allows to reconstruct the phase of the THz waveform from the measured envelope and is a generalization of the so-called Kramers-Kronig (KK) receiver in optical communications[1,2]. In our experiments, we use a high-speed Schottky-barrier diode (SBD) as a broadband and compact envelope detector. In contrast to a conventional photodetector used in optical communications, the SBD features non-quadratic rectification characteristics, which must be accounted for in our newly proposed generalized KK algorithm. We demonstrate transmission of quadrature phase-shift keying (QPSK) and 16QAM signals at a carrier frequency of 0.3 THz over a distance of 110 m, achieving line rates of up to 132 Gbit/s and net data rates of up to 115 Gbit/s after subtraction of the error correction overhead. To the best of our knowledge, this is the first experiment using an envelope detector and subsequent KK processing in wireless communications, leading to the highest data rate so far demonstrated for wireless THz transmission over distances of more than 100 m.

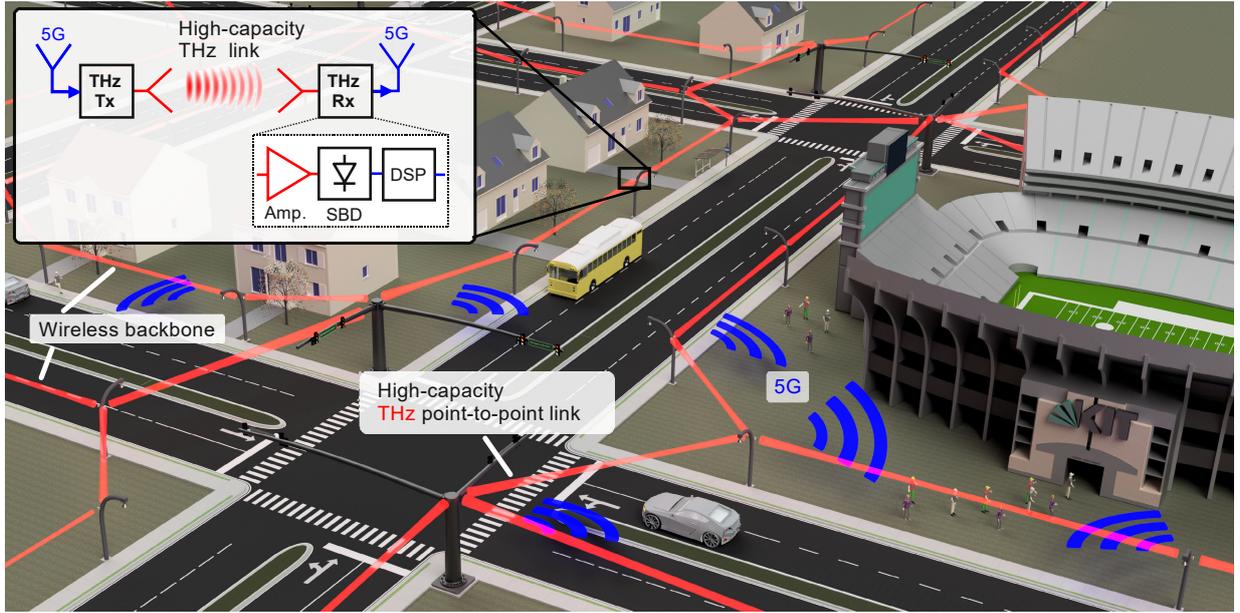

**Fig. 1: Vision of a future wireless backbone network.** High-capacity THz line-of-sight links connect small 5G radio cells with a coverage of only a few tens of meters. Each of these cells locally guarantees ultra-broadband wireless services. The wireless connections allow a flexible and efficient installation without the need for deploying optical fibres or changing fibre installations in case reconfigurations are required. The inset shows a schematic of a THz link based on a Kramers-Kronig-(KK-) type receiver. This scheme greatly reduces the complexity of the receiver hardware by replacing technically complex THz LOs and mixer circuits with a simple Schottky-barrier diode (SBD).

The vision of a future wireless backbone network is shown in Fig. 1. Small 5$^{\text{th}}$-Generation (5G) radio cells[21,22] with a coverage of only a few tens of meters enable frequency reuse and guarantee broadband service for large numbers of terminal devices. The various cells are connected by high-capacity THz line-of-sight links in a mesh configuration, allowing to increase network resilience and flexibility while decreasing installation costs in comparison to fibre-based connections. The inset shows a schematic of the THz link using a KK receiver scheme. The KK approach greatly reduces the complexity of the receiver hardware, which only consists of an amplifier and a Schottky-barrier diode (SBD) and does not require any technically complex THz LOs and mixer circuits.

On a fundamental level, KK processing[1] relies on the fact that the real part and the imaginary part of an analytic time-domain signal represent a Hilbert transform pair. This relationship can be translated into an equivalent relationship that allows to retrieve the phase of a complex signal once its amplitude has been measured, see Methods for details. For the KK-based reception described here, the THz signal incident at the receiver can be described as

$$u_{\text{THz}}(t) = \Re\left\{ \underline{U}(t) e^{j2\pi f_{\text{THz}} t} \right\}, \quad (1)$$

where $\Re\{.\}$ denotes the real part and where the complex envelope $\underline{U}(t)$ consists of a strong real-valued constant part $U_0$ and an analytic data signal $\underline{U}_s(t)$ with single-sided power spectrum,

$$\underline{U}(t) = U_0 + \underline{U}_s(t) = |\underline{U}(t)| e^{j\Phi(t)}. \quad (2)$$

To ensure correct reconstruction of the phase $\Phi(t)$ from the measured amplitude $|\underline{U}(t)|$, $\underline{U}(t)$ must be a minimum-phase signal in the time domain, which is ensured by[1]

$$|\underline{U}_s(t)| < U_0 \quad \text{for all } t. \quad (3)$$

The phase $\Phi(t)$ can then be reconstructed by a KK-type relation,

$$\Phi(t) = \frac{1}{\pi} \mathcal{P} \int_{-\infty}^{\infty} \frac{\ln(|\underline{U}(t)|)}{t-\tau} d\tau, \quad (4)$$

where $\mathcal{P}$ stands for the Cauchy principal value of the otherwise undefined improper integral. A more detailed derivation can be found in the Methods.

In optical communications, a photodiode is commonly used as an envelope detector for KK reception. Since the photocurrent is proportional to the incident optical power, the magnitude of the optical amplitude can be digitally reconstructed from the electrical signal by a simple square-root operation. In contrast to that, SBD feature complex rectification characteristics. This requires a generalization of the KK-based signal reconstruction algorithm. To this end, we assume that the relationship between the amplitude $|\underline{U}(t)|$ of the THz signal at the SBD receiver input and the current $i(t)$ at the output can be described by a bijective function $g$,

$$i(t) = g(|\underline{U}(t)|), \qquad (5)$$

see Supplementary Section S2 for details. The amplitude $|\underline{U}(t)|$ can then be reconstructed by applying the inverse function $g^{-1}(i)$ to the measured output current, and the phase $\Phi(t)$ can be found by replacing $\ln(|\underline{U}(t)|)$ in Eq. (4) with the generalized expression $\ln(g^{-1}(i))$,

$$\Phi(t) = \frac{1}{\pi}\mathcal{P}\int_{-\infty}^{\infty}\frac{\ln(g^{-1}(i))}{t-\tau}\,d\tau. \qquad (6)$$

In contrast to square-law photodetectors, the exact rectification characteristics of SBDs depend crucially on the operating point as well as on the peripheral THz and baseband circuits and are hence hard to describe analytically. We therefore express the inverse function $g^{-1}(i)$ as a power series with $N+1$ initially unknown coefficients $a_n$,

$$|\underline{U}_R| = g^{-1}(i) = \sum_{n=0}^{N} a_n i^n. \qquad (7)$$

In this relation, $|\underline{U}_R|$ denotes the THz voltage amplitude that is reconstructed from a measured current $i$ at the SBD output. The coefficients $a_n$ have to be determined from measured data. To this end, we use test signals with known amplitude $|\underline{U}(t)|$ and measure the associated current $i(t)$ at discrete times $t_1, t_2, \ldots t_M$ with $M \gg N$. The coefficients $a_n$ are obtained from a least-squares fit of the polynomial function $|\underline{U}_R(t_m)| = g^{-1}(i(t_m))$ given by Eq. (7) to the known data set $|\underline{U}(t_m)|$, see Supplementary Section S3 for details. The result is shown in Fig. 2, together with the square-root characteristics that would be used in conventional KK processing of optical signals. While the square-root characteristics can be used as an approximation for small signal amplitudes, the generalized approach accounts for the actual behaviour of the SBD receiver over a large range of signal levels. In addition, the generalized function comprises an offset $a_0 \neq 0$ that assigns a zero output current $i = 0$ to a non-zero THz amplitude $|\underline{U}_R| = g^{-1}(0)$, see Supplementary Section S3. This offset avoids small values of the reconstructed THz amplitude, which would otherwise require artificial clipping[23] of the signals at low reconstructed voltages $|\underline{U}_R|$ to mitigate large uncertainties in the phase reconstruction according to Eq. (4) due to the singularity of $\ln(|\underline{U}_R|)$ at $|\underline{U}_R| = 0$. Note that the measurement data used for the least-squares fit were obtained from an SBD in combination with a THz amplifier at its input. The extracted receiver characteristics $g^{-1}(i)$ hence account not only for the SBD but can also compensate the saturation behaviour of the THz amplifier. More details on the characterization of the SBD and the THz amplifier can be found in Supplementary Sections S3 and S4. Note also that the non-quadratic characteristics of the SBD can broaden the received signal spectrum and might increase the bandwidth requirements of the receiver circuits. This effect, however, is not very prominent, see Supplementary Section S5.

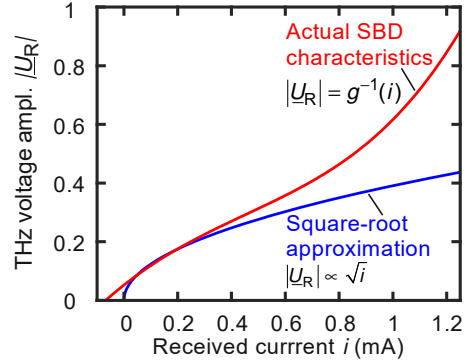

**Fig. 2: SBD receiver characteristics.** The SBD rectifies the incident THz signal leading to an output current $i$ that depends on the THz voltage amplitude $|\underline{U}|$. We describe the actual receiver characteristics $|\underline{U}_R| = g^{-1}(i)$ (red) by a power-series approximation according to Eq. (7). The coefficients $a_n$ are determined from measured test signals, see Supplementary Section S3. As a reference, we show the square-root relationship (blue) that is assumed in conventional Kramers-Kronig processing (conv. KK) and that leads to a relationship of the form $|\underline{U}_R| \propto \sqrt{i}$. The generalized approach describes the actual behaviour of the SBD receiver for all signal levels, whereas the square-root characteristics can be used as an approximation for small signal levels only. Note that the generalized function comprises an offset $a_0 \neq 0$ that assigns a non-zero THz amplitude $|\underline{U}_R| = g^{-1}(0)$ to a zero output current $i = 0$, thereby effectively clipping small signal amplitudes[23], see Supplementary Section S3 for details.

To demonstrate the viability of generalized KK processing, we perform wireless transmission experiments at a carrier frequency of $f_{\text{THz}} = 0.3\,\text{THz}$. Figure 3 depicts a simplified sketch of the experimental setup; a comprehensive description is given in Supplementary Section S1. At the transmitter (Tx), the THz signal is generated by frequency down-conversion of an optical signal via photomixing. An arbitrary-waveform generator (AWG) is used to drive an electro-optic IQ-modulator, which encodes data signals with bandwidth $B$ on an optical carrier with frequency $f_0$. The optical data signal is combined with two continuous-wave (c.w.) tones at optical frequencies $f_1$ and $f_2$ and then fed to a high-speed uni-travelling-carrier photodiode (UTC-PD)[24] for photomixing. The optical spectrum at the input of the photodiode is shown in Inset 1 of Fig. 3. The first optical c.w. tone at frequency $f_1$ serves as an LO for KK reception, whereas the second c.w. tone at $f_2$ is used for down-converting the data signal and the LO tone to the THz carrier frequency $f_{THz} = f_1 - f_2$. The LO tone at frequency $f_1$ is placed at the edge of the data signal spectrum, and the data signal can hence be represented by a complex analytic signal $\underline{U}_s(t)$ with respect to the carrier frequency $f_{THz}$. The total voltage envelope is then given by $\underline{U}(t) = U_0 + \underline{U}_s(t)$, where $U_0$ denotes the LO amplitude generated by the optical tone at $f_1$. For distortion-free KK processing, the power of the c.w. LO tone has to be chosen large enough to guarantee that the resulting complex envelope $\underline{U}(t)$ fulfils the minimum-phase condition according to Eq. (3). Note that, in contrast to conventional SBD-based self-coherent transmission[25,26], generalized KK processing avoids unused guard bands between the LO tone and the payload signal that would occupy at least half of the available transmission bandwidth.

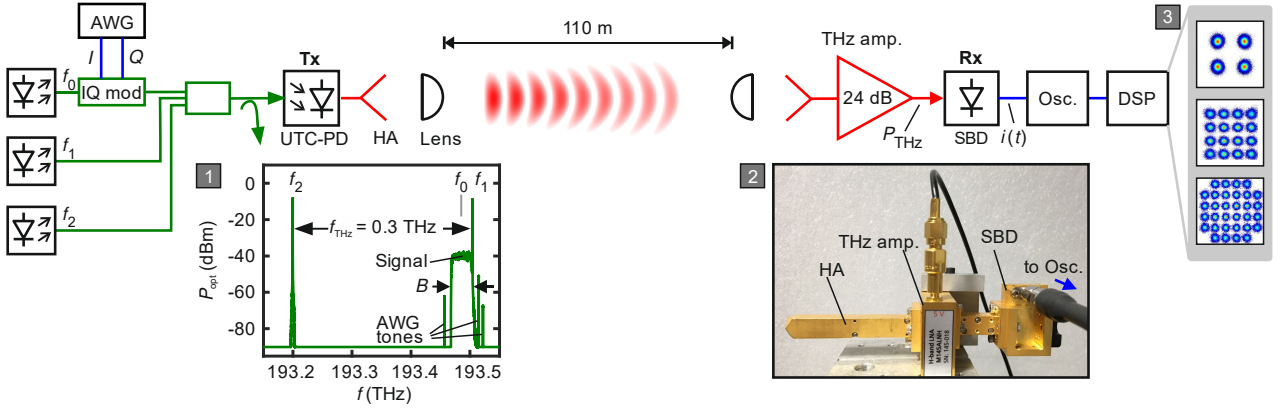

**Fig. 3: Experimental setup.** In a first step, an optical data signal is generated at carrier frequency $f_0$ by an in-phase/quadrature (I/Q) modulator that is driven by an arbitrary waveform generator (AWG). The optical data signal is then combined with two continuous-wave (c.w.) tones at optical frequencies $f_1$ and $f_2$. The tone at frequency $f_1$ is placed at the edge of the data spectrum and will eventually act as an LO for KK reception, whereas the second tone at frequency $f_2$ serves as a reference tone for down-converting the data signal to the THz carrier frequency $f_{THz} = f_1 - f_2$ by photomixing in a high-speed uni-travelling-carrier photodiode (UTC-PD). Inset 1 shows the corresponding optical spectrum (resolution bandwidth = 180 MHz) recorded after a monitoring tap. The optical spectrum also exhibits spurious tones generated by the AWG ("AWG tones"). The overall optical power entering the UTC-PD is between 10 dBm and 14 dBm, depending on the desired THz power. At the output of the UTC-PD, the THz signals are radiated into free space by a horn antenna (HA) and a subsequent polytetrafluoroethylene (PTFE) collimation lens. After a transmission distance of 110 m, the THz signal is received by a second lens and another horn antenna and fed to a THz amplifier providing a 24 dB gain. The Schottky barrier diode (SBD) is connected to the output of the amplifier, and a real-time oscilloscope (Osc.) is used to capture the output current of the SBD for further off-line processing. Inset 2 shows a photograph of the receiver including the horn antenna, the THz amplifier and the SBD. Exemplary constellation diagrams for QPSK, 16QAM and 32QAM are shown in Inset 3 for symbol rates of 30 GBd, 15 GBd and 5 GBd, respectively.

After photomixing, the THz signals are radiated into free space by a horn antenna (HA). A subsequent polytetrafluoroethylene (PTFE) lens collimates the beam. After a transmission distance of 110 m, the THz signal is collected by a second lens and another horn antenna. The transmission loss of the free-space section amounts to 19 dB and is (over-)compensated by a low-noise amplifier with 24 dB gain. The SBD is connected to the output of the amplifier, see Inset 2 of Fig. 3, and a real-time oscilloscope with 80 GHz bandwidth is used to record the output current $i(t)$. After data acquisition, we apply the generalized KK algorithm, followed by blind coherent digital signal processing, see Methods. This leads to the constellation diagrams shown in Inset 3 of Fig. 3.

In our experiments, we explore QPSK, 16QAM, and 32QAM as modulation formats. The results of QPSK and 16QAM are summarized in Fig. 4. We use generalized KK processing as described above (gen. KK) and compare it to the results obtained by conventional KK processing when assuming a square-root relationship between the SBD output current and the THz amplitude (conv. KK). We also consider evaluation assuming heterodyne reception without any KK processing (w/o KK). In this case, down-conversion leads to an LO-signal mixing product which is impaired by the nonlinear mixing of the signal with itself, see Methods for details. In a first series of experiments, we analyse the influence of the optical carrier-to-signal power ratio (CSPR) on the signal quality for QPSK modulation, see Fig. 4a. The CSPR refers to the power ratio between the LO tone and the data signal,

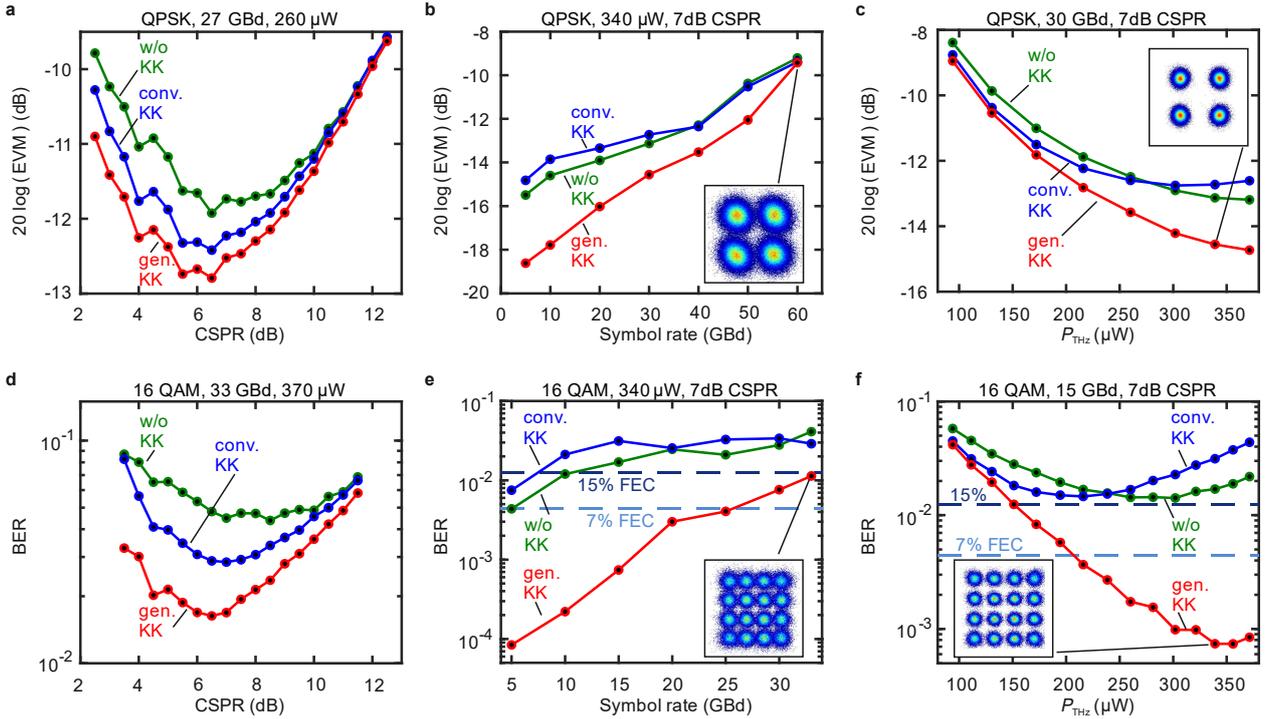

**Fig. 4: Results of the transmission experiments using QPSK and 16QAM signals, all transmitted over a free-space distance of 110 m. a,** Error vector magnitude (EVM) as a function of the carrier-to-signal power ratio (CSPR) for QPSK signals measured at a symbol rate of 27 GBd for a constant incident THz power of $P_{THz}$ = 260 µW. The CSPR refers to the power ratio between the LO tone and the data signal. For generalized Kramers-Kronig processing ("gen. KK"), the actual SBD receiver characteristics are taken into account and are compensated, see Eqs. (6) and (7), leading to the best transmission performance. As a reference, we consider conventional KK reception ("conv. KK") assuming a quadratic relationship between the SBD current $i$ and the THz voltage amplitude $|U|$. The curve "w/o KK" indicates heterodyne reception without any KK processing and without a guard band between the signal and the LO. This scheme suffers from impairments due to nonlinear interaction of the signal with itself. For all processing schemes, an optimum CSPR of around 6…7 dB is found, representing an ideal trade-off between low signal power at high CSPR and low LO power at small CSPR. **b,** EVM as a function of the symbol rate for a CSPR of 7 dB and a total incident THz power of $P_{THz}$ = 340 µW. The inset shows the constellation diagram for a symbol rate of 60 GBd, for which an EVM of -9.4 dB is recorded. For this case, also a bit error ratio (BER) of $1.9 \times 10^{-3}$ could be measured, which is below the threshold for forward-error correction (FEC) with 7 % FEC overhead. A net data rate of 112 Gbit/s results. **c,** EVM as a function of the incident THz power $P_{THz}$ for a symbol rate of 30 GBd and a CSPR of 7 dB. The performance advantages of generalized KK processing are most pronounced at high THz powers, where the SBD receiver characteristics cannot be approximated by a square-root relationship. **d,** BER as a function of the CSPR for 16QAM transmission at a symbol rate of 33 GBd at an incident THz power of $P_{THz}$ = 370 µW. The CSPR analysis exhibits a similar behaviour as for QPSK, showing best performance for a CSPR of approximately 6…7 dB. **e,** BER as a function of the symbol rate with a CSPR of 7 dB and an incident THz power of $P_{THz}$ = 340 µW. Employing the generalized KK reception, the BER stays below the 15 % FEC limit and leads to a line rate of 132 Gbit/s and a net date rate of 115 Gbit/s when accounting for FEC overhead. This is the highest value so far achieved for wireless THz transmission over distances of more than 100 m. The constellation diagram for a symbol rate of 33 GBd is shown in the inset. **f,** BER as a function of the incident THz power $P_{THz}$ for a symbol rate of 15 GBd and a CSPR of 7 dB. For high THz powers, generalized KK processing allows to reduce the BER by more than an order of magnitude.

$$\mathrm{CSPR} = 10\log\left(|U_0|^2 \Big/ \overline{|U_s(t)|^2}\right), \qquad (8)$$

where $\overline{|U_s(t)|^2}$ denotes the time-average of the squared signal voltage. When changing the CSPR, the sum of carrier and signal power is kept constant at $P_{THz} = 260\,\mu\mathrm{W}$, measured at the input port of the SBD. The high signal quality of the received QPSK constellation diagram leads to a small number of errors for the $3\times10^5$ evaluated symbols, and the measured BER does not represent a reliable estimate for the bit error probability. We therefore use the error vector magnitude (EVM) as a quality metric[27], see Methods for details. We find an optimum CSPR value of approximately 6…7 dB for both types of KK processing as well as for heterodyne reception without KK processing. For smaller CSPR, the EVM obtained from KK processing increases, because for a constant $P_{THz}$ the LO magnitude becomes smaller, and the minimum-phase condition according to Eq. (3) is violated. For heterodyne reception without KK processing, the nonlinear interaction of the signal with itself can no longer be neglected if the CSPR is too small, thus impairing the signal quality. For larger CSPR and a constant $P_{THz}$, the data signal becomes smaller and is finally dominated by receiver noise.

Next, we evaluate the EVM for various symbol rates of up to 60 GBd at a total THz power of $P_{THz} = 340$ µW and for a CSPR of 7 dB, Fig. 4b. For lower symbol rates, generalized KK processing improves the signal quality. At higher symbol rates, thermal noise and quantization noise are the dominant limitations, and the three processing schemes converge in their performance. The inset of Fig. 4b shows the constellation diagram for the 60 GBd transmission, for which an EVM of – 9.4 dB is measured. For this case, the bit error probability could be reliably estimated by measuring a bit error ratio (BER). The BER of $1.9 \times 10^{-3}$ is below the threshold for forward error correction (FEC) with 7 % overhead[28] ($4.4 \times 10^{-3}$) and leads to a net data rate of 112 Gbit/s.

In a third measurement, we investigate the dependence of the EVM on the incident THz power for a symbol rate of 30 GBd and a CSPR of 7 dB, Fig. 4c. For small THz powers, where the SBD characteristics can be approximated by a square-root relationship, the performance of generalized KK processing is similar to that of the conventional KK scheme. In this regime, KK processing does not show a significant advantage over heterodyne reception, because the unwanted interaction products of the signal with itself are comparable to or even smaller than the receiver noise power. At large THz powers, we observe an improvement of the EVM from -12.6 dB obtained from conventional KK processing to -14.7 dB for the generalized KK scheme. Assuming that the signal impairments can be modelled as additive white Gaussian noise, this would correspond to a BER improvement[27] from $1 \times 10^{-5}$ to $3 \times 10^{-8}$.

The same set of measurements is repeated with 16QAM signalling, i.e., at double the spectral efficiency. In these experiments, we were able to directly measure the BER for a sequence of $1.6 \times 10^5$ received symbols. The CSPR analysis, Fig. 4d, exhibits a similar behaviour as for QPSK, showing best performance, i.e., minimum BER, for a CSPR of approximately $6 \ldots 7$ dB. The evaluation of the BER for various symbol rates at a total THz power of $P_{THz} = 340$ µW and for a CSPR of 7 dB is shown in Fig. 4e. When using heterodyne reception without KK processing, the BER is larger than the FEC-limit assuming a 7 % overhead, even for symbol rates as small as 10 GBd. Conventional KK processing does not significantly improve the BER, because higher-order mixing products in the SBD have noticeable impact for the comparatively high THz power of 340 µW. In contrast to this, generalized KK processing allows to significantly improve the BER, reaching a reduction by more than an order of magnitude for small symbol rates. For higher symbol rates, thermal noise limits again the performance of the reception, and the advantage of generalized KK processing becomes smaller. Nevertheless, for a 16QAM symbol rate of 33 GBd and generalized KK processing, the BER is still below the 15 % FEC limit[29] of $1.25 \times 10^{-2}$. This leads to a line rate of 132 Gbit/s before FEC and to a net data rate of 115 Gbit/s. To the best of our knowledge, this represents the highest data rate so far demonstrated for wireless THz transmission over practically relevant distances of more than 100 m.

Finally, we investigate the dependence of the BER on the incident THz power for a symbol rate of 15 GBd and a CSPR of 7 dB, Fig. 4f. We find that generalized KK processing leads to a clear performance advantage over a wide range of incident THz powers. This demonstrates the capability of our approach, especially if phase-noise sensitive spectrally efficient quadrature-amplitude modulation formats are used. We also tested 32QAM modulation, observing similar performance advantages of the generalized KK scheme, but lower overall data rates, see Supplementary Section S6.

In summary, we have demonstrated a greatly simplified coherent receiver scheme for THz data signals combining a Schottky-barrier diode (SBD) as simple envelope detector with generalized Kramers-Kronig signal processing. Our scheme accounts for the real characteristics of the SBD and leads to a significant performance improvement compared to conventional square-root Kramers-Kronig processing or to heterodyne reception without a guard band. We demonstrate the viability of the scheme using QPSK, 16QAM, and 32QAM signalling and achieve line rates of up to 132 Gbit/s, which corresponds to net data rates of up to 115 Gbit/s when accounting for the overhead of forward error correction (FEC). To the best of our knowledge, this represents the highest data rate so far demonstrated for wireless THz transmission over practically relevant distances of more than 100 m.

## Acknowledgements

This work was supported by the European Research Council (ERC Consolidator Grant 'TeraSHAPE', # 773248), by the Alfried Krupp von Bohlen und Halbach Foundation, by the Helmholtz International Research School of Teratronics (HIRST), and by the Karlsruhe School of Optics and Photonics (KSOP). The work relies on instrumentation funded by the European Regional Development Fund (ERDF, Grant EFRE/FEIH_776267), by the Deutsche Forschungsgemeinschaft (DFG; Grants DFG/INST 121384/166-1 and DFG/INST 121384/167-1), and by the Hector Stiftung.

## Methods

**Schottky-barrier diode.** We use a zero-bias Schottky-barrier diode[30,31] (SBD, Virginia Diodes Inc., model WR3.4ZBD-F) as THz envelope detector. This device offers high responsivity of the order of 2000 V/W along with broadband output circuitry (bandwidth approx. 40 GHz) for down-conversion of high-speed data signals. The WR 3.4 input port of the SBD allows to directly connect the SBD to the output of a THz low-noise amplifier[32], designed for operation in the submillimetre H-band (0.220 THz - 0.325 THz), see Supplementary Section S4 for details.

**Phase reconstruction by Kramers-Kronig (KK) processing:** On a fundamental level, KK processing[1] relies on data signals that are analytic. For an arbitrary analytic time-domain signal $\underline{s}(t) = s_r(t) + j s_i(t)$, the real part and the imaginary part are connected by a Hilbert transform,

$$s_i(t) = \frac{1}{\pi} \mathcal{P} \int_{-\infty}^{\infty} \frac{s_r(\tau)}{t-\tau} d\tau = \mathcal{P}\left\{\frac{1}{\pi t} * s_r(t)\right\}, \quad (9)$$

$$s_r(t) = -\frac{1}{\pi} \mathcal{P} \int_{-\infty}^{\infty} \frac{s_i(\tau)}{t-\tau} d\tau = -\mathcal{P}\left\{\frac{1}{\pi t} * s_i(t)\right\}, \quad (10)$$

where the symbol $*$ denotes a convolution, and where the operator $\mathcal{P}$ stands for the Cauchy principal value of the subsequent improper integral. Equations (9) and (10) can be interpreted as the time-domain analogon of the KK relations[5,6] that connect the real and the imaginary part of a transfer function belonging to a system that is causal in the time domain[33].

Analytic signals as defined by Eqs. (9) and (10) feature single-sided power spectra, which do not contain any spectral components for negative frequencies $\omega < 0$. Note that the mutual relationships between the imaginary and the real part of an analytic signal are only defined if both the real and the imaginary part are zero-mean. To understand this, let us consider a signal with real part $s'_r(t) = s_r(t) + C_r$ that consists of a constant DC part $C_r$ and a zero-mean signal $s_r(t)$. When introducing $s'_r(t)$ into Eq. (9), the integral has to be evaluated in the sense of a Cauchy principal value to ensure convergence both for the singularity at $t = \tau$ and for the infinite integration limits,

$$s_i(t) = \frac{1}{\pi} \lim_{\varepsilon \to 0} \left( \int_{t-1/\varepsilon}^{t-\varepsilon} \frac{s'_r(\tau)}{t-\tau} d\tau + \int_{t+\varepsilon}^{t+1/\varepsilon} \frac{s'_r(\tau)}{t-\tau} d\tau \right). \quad (11)$$

In this case, the constant $C_r$ does not contribute to the imaginary part $s_i(t)$, and when using Eq. (9) to reconstruct the real part, one would obtain $s_r(t)$ rather than $s'_r(t)$. Similarly, using Eq. (9) to reconstruct the real part $s_r(t)$ from an imaginary part $s'_i(t) = s_i(t) + C_i$ with non-zero mean value $C_i$ would suppress the DC part $C_i$ and produce the real part $s_r(t)$ that belongs to the corresponding zero-mean signal $s_i(t)$. Hence, signals with non-zero mean value, i.e., non-zero spectral power at $\omega = 0$, are no analytic signals in the sense of Eqs. (9) and (10).

For KK-based transmission, the analytic data signal $\underline{U}_s(t)$ is superimposed with a carrier tone, which is represented by a constant voltage $U_0$ in the baseband. Without loss of generality, we assume $U_0$ to be real-valued and positive. The overall complex envelope $\underline{U}(t)$ can then be written as

$$\underline{U}(t) = U_0 + \underline{U}_s(t) = |\underline{U}(t)| e^{j\Phi(t)}, \quad (12)$$

where $|\underline{U}(t)|$ and $\Phi(t)$ denote the amplitude and the phase of the overall complex envelope $\underline{U}(t)$, comprising the data signal $\underline{U}_s(t)$ and the carrier amplitude $U_0$.

In technical implementations of KK reception, the amplitude $|\underline{U}(t)|$ can be directly measured through some sort of envelope detector with nonlinear characteristics such as a photodiode or, as in our case, an SBD. To establish a relation between the amplitude $|\underline{U}(t)|$ and the phase $\Phi(t)$ of the complex envelope $\underline{U}(t)$, we derive an auxiliary signal $\underline{s}(t)$ by applying the complex natural logarithm to Eq. (12),

$$\underline{s}(t) = \ln(\underline{U}(t)) = \ln(|\underline{U}(t)|) + j\Phi(t). \quad (13)$$

For the further analysis, we exploit the fact that $\underline{s}(t)$ is an analytic signal if the complex data signal $\underline{U}_s(t)$ is an analytic signal, too, and if in addition the overall complex envelope $\underline{U}(t)$ is minimum-phase[1]. A necessary condition for $\underline{U}(t)$ being minimum-phase is that the associated time-dependent trajectory described by $\underline{U}(t)$ in the complex plane does not encircle the origin, which is ensured if $|\underline{U}_s(t)| < U_0, \forall t$. In this case, we may apply Eq. (9) to the auxiliary analytic signal $\underline{s}(t)$ and obtain

$$\Phi(t) = \frac{1}{\pi} \mathcal{P} \int_{-\infty}^{\infty} \frac{\ln(|\underline{U}(t)|)}{t-\tau} d\tau. \quad (14)$$

This relation allows to reconstruct the phase $\Phi(t)$ of $\underline{U}(t)$ if only the amplitude $|\underline{U}(t)|$ is known. The complex data signal $\underline{U}_s(t) = |\underline{U}_s(t)| \exp(j\varphi_s(t))$ is then recovered from $\underline{U}(t)$ by subtracting the DC voltage $U_0$ corresponding to the amplitude of the carrier tone, see Eq. (12),

$$\underline{U}_s(t) = |\underline{U}_s(t)| e^{j\varphi_s(t)} = |\underline{U}(t)| e^{j\Phi(t)} - U_0 \quad (15)$$

Note that the voltage magnitude $|\underline{U}(t)|$ in Eq (14) can be scaled with an arbitrary constant factor $\zeta$ without changing the reconstructed phase $\Phi(t)$. This becomes plausible when considering that $\ln(\zeta |\underline{U}(t)|) = \ln(|\underline{U}(t)|) + \ln(\zeta)$, and that the constant $\ln(\zeta)$ is suppressed when applying the Hilbert transform according to Eq. (14), see discussion after Eq. (11).

**Heterodyne reception without guard band.** In our evaluation, we also consider single-ended heterodyne reception without any KK processing (w/o KK), which does not account for nonlinear interaction of the signal with itself. According to Eq. (5), the SBD current $i(t)$ depends on the magnitude of the complex envelope $|\underline{U}(t)|$. For the sake of simplicity, we assume a square-law relationship $i(t) \propto |\underline{U}(t)|^2$ in the following derivation. With Eq. (2) we obtain

$$i(t) \propto |U_0 + \underline{U}_s(t)|^2 = U_0^2 + |\underline{U}_s(t)|^2 + 2U_0 \Re(\underline{U}_s(t)), \quad (16)$$

where $\Re\{\cdot\}$ denotes the real part. The component $2U_0 \Re\{\underline{U}_s(t)\}$ corresponds to the mixing product of the LO tone with amplitude $U_0$ and the analytic data signal $\underline{U}_s(t)$ and allows to recover the data signal. However, there is also the mixing product $|\underline{U}_s(t)|^2$ of the signal with itself, which covers the same spectral region as the mixing product $2U_0 \Re(\underline{U}_s(t))$ unless a guard band between $U_0$ and the data spectrum is used. This mixing product degrades the received signal quality and can be compensated by using KK processing as described in the main text. For large carrier-to-signal power ratios (CSPR), i.e., large values of $U_0$, the data can still be recovered without using KK processing, because the data signal $2U_0 \Re(\underline{U}_s(t))$ dominates over the unwanted mixing product for $|U_0| \gg |\underline{U}_s(t)|$.

**Digital signal processing chain.** In our experiments, we digitize the receiver currents using a real-time oscilloscope (Keysight UXR0804A) and store them for subsequent offline digital signal processing (DSP) and signal quality evaluation. The transmitted data signal $|\underline{U}_s(t)| \exp(j\varphi_s(t))$ is reconstructed from the digitally captured SBD current $i$ by generalized KK processing, Eq. (5) and (6). We first calculate the THz voltage amplitude $|\underline{U}(t)|$ from the captured signal by inverting Eq. (5). For the phase reconstruction, Eq. (6), the Hilbert transform is implemented as a convolution with a discrete-time finite-impulse-response filter[34]. For the processing, an oversampling of 6 samples per symbol is used to avoid performance penalties due to spectral broadening[35] introduced by the natural logarithm in Eq. (6) and by the function Eq. (7). After phase reconstruction, the complex signal $\underline{U}(t)$ is down-sampled to a ratio of 2 samples per symbol, and the analytic signal $\underline{U}_s(t)$ is obtained according to Eq. (15). The original QAM signal is then recovered by down-converting $\underline{U}_s(t)$ to the baseband. The KK receiver DSP is followed by the DSP blocks of a fully blind coherent optical receiver[36] with the exception that only a single polarization is evaluated. Specifically, we perform a timing-error estimation in the frequency domain, followed by a compensation with a Lagrange interpolator in a Farrow[37] structure. We then apply an adaptive feed-forward equalizer (FFE), whose coefficients are adapted blindly using the constant-modulus algorithm[38] (CMA). The remaining frequency offset originating from frequency drifts of the LO laser at $f_1$ as well as of the signal laser at $f_0$ is removed, and a subsequent phase recovery using the blind phase search algorithm[39] combats laser phase noise. Finally, we apply a real-valued multiple-input multiple-output (MIMO) FFE with the in-phase and quadrature component of the complex baseband as inputs to compensate IQ imbalances of the transmitter hardware[40]. The FFE coefficients are adapted by a decision-directed least-mean square algorithm[41]. Finally, the bit error ratio (BER) and the error vector magnitude (EVM) are calculated.

**Error vector magnitude (EVM) and bit error ratio (BER).** The EVM and BER are standard metrics to evaluate the quality of communication signals and are widely used in the optical as well as in the THz communications community[27]. Assuming data-aided reception of $N$ randomly transmitted symbols, the EVM is a measure for the effective distance of the received complex symbols $E_r$ from their ideal positions $E_t$ in the constellation diagram. Specifically, it relates the root mean square $\sigma_{err}$ of the error vector amplitude $|E_r - E_t|$ to the average amplitude $|E_{t,a}|$ of the $N$ ideal constellation points,

$$\mathrm{EVM} = \frac{\sigma_{err}}{|E_{t,a}|}, \quad \mathrm{EVM}_{dB} = 20\log(\mathrm{EVM}),$$
$$|E_{t,a}|^2 = \frac{1}{N}\sum_{i=1}^{N}|E_{t,i}|^2, \quad \sigma_{err}^2 = \frac{1}{N}\sum_{i=1}^{N}|E_{r,i} - E_{t,i}|^2. \quad (17)$$

The BER is obtained by comparing the digitally processed bit sequence with the known transmitter bit sequence, and by relating the number of counted errors to the total number of bits. For the results shown in Fig. 4, a random symbol pattern of length $2^{15}$ (QPSK) or $2^{14}$ (16QAM) is periodically repeated. The real-time oscilloscope records a sequence of $8\times10^6$ samples at a sampling rate of 128 GSa/s for symbol rates smaller than 20 GBd, and at a sampling rate of 256 GSa/s for symbol rates $\geq 20\,\mathrm{GBd}$. This corresponds to recording lengths of 31.25 µs, and 62.5 µs, respectively. The BER and EVM evaluation is restricted to the last 10 transmit pattern periods to ensure that the equalizers already converged. In case of QPSK modulation, the signal quality is so good that only very few errors can be found for the given recording length. Generally, if less than 13 errors are counted in a recording, the measured BER is not a reliable estimate[42] of the actual bit error probability due to poor statistics. Moreover, since the constellation points seen in Fig. 4 are not perfectly circular, the assumption of additive white Gaussian noise is violated, and we refrain from estimating the BER from the EVM. Therefore, the measured EVM itself is used as a quality metric for QPSK signals.

# Generalized Kramers-Kronig Receiver for Coherent THz Communications (Supplementary Information)

T. Harter[1,2*], C. Füllner[1*], J. N. Kemal[1], S. Ummethala[1,2], J. L. Steinmann[3], M. Brosi[3],
J. L. Hesler[4], E. Bründermann[3], A.-S. Müller[3], W. Freude[1], S. Randel[1**], C. Koos[1,2**]

[1]*Institute of Photonics and Quantum Electronics (IPQ), Karlsruhe Institute of Technology (KIT), Germany*
[2]*Institute of Microstructure Technology (IMT), Karlsruhe Institute of Technology (KIT), Germany*
[3]*Institute for Beam Physics and Technology (IBPT), Karlsruhe Institute of Technology (KIT), Germany*
[4]*Virginia Diodes, Inc. (VDI), Charlottesville, United States of America*

*These authors contributed equally to this work
**e-mail: sebastian.randel@kit.edu, christian.koos@kit.edu

## S1. Experimental setup

The experimental setup is depicted in Fig. S1. An arbitrary-waveform generator (AWG, Keysight M8194A) is used to drive an IQ-modulator to encode a quadrature-amplitude modulated signal onto an optical carrier at frequency $f_0$. The data pulses have a raised-cosine spectrum with bandwidth $B$ and roll-off factor 0.1. The modulated optical signal is amplified by an erbium-doped fibre amplifier (EDFA), followed by a 0.6 nm optical bandpass filter to suppress amplified spontaneous emission (ASE) noise. A 50/50-coupler combines the modulated optical carrier with an unmodulated optical continuous-wave

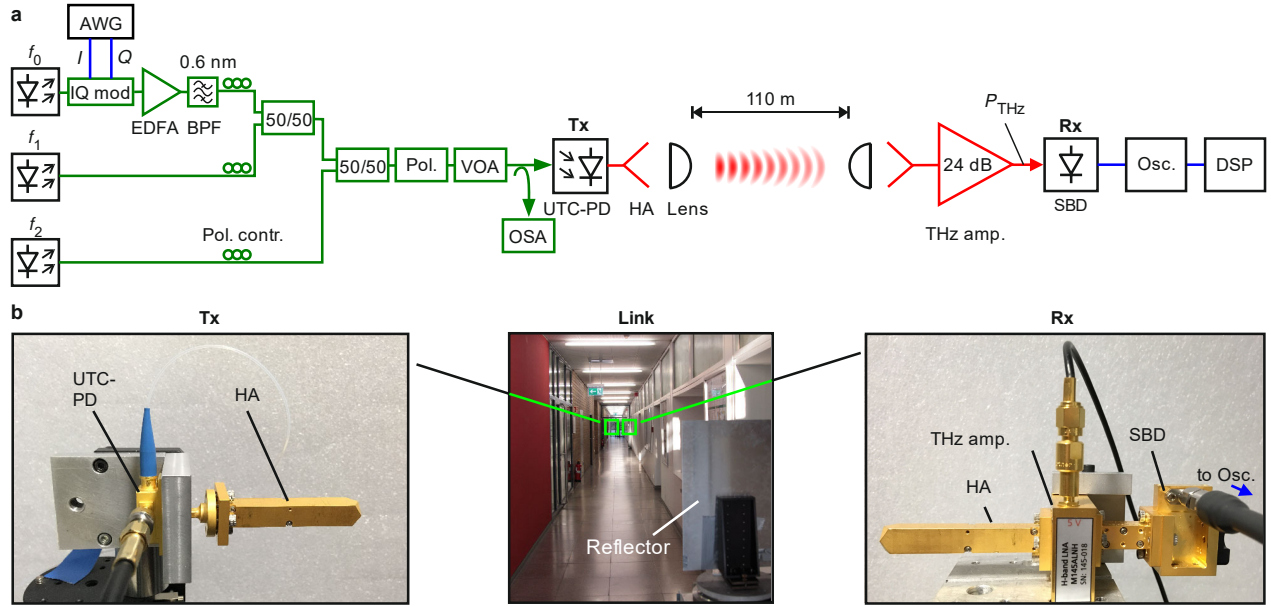

**Fig. S1: Experimental setup.** The data signal is modulated on an optical carrier at frequency $f_0$. The optical data signal is superimposed with an unmodulated optical tone $f_1$, which, after down-conversion, will form the THz local oscillator (LO) for generalized Kramers-Kronig reception. The optical data signal and the LO are converted to the THz domain by photomixing with an unmodulated carrier $f_2 = f_1 - f_{THz}$ in a high-speed uni-travelling carrier photodiode (UTC-PD), see Inset 1 of Fig. 3 of the main manuscript for an optical spectrum measured at the input of the UTC-PD. Horn antenna (HA) / lens combinations are used to radiate and receive the THz signals. The transmission distance amounts to 110 m. After amplification by 24 dB, a Schottky-barrier diode (SBD) rectifies the THz signal. A real-time oscilloscope (Osc.) is used to capture the RF signal at the SBD output for off-line processing. AWG: arbitrary waveform generator. IQ mod: in-phase and quadrature modulator. EDFA: Erbium-doped fibre amplifier. BPF: optical bandpass filter. Pol.: polarizer. VOA: variable optical attenuator. OSA: optical spectrum analyser. HA: horn antenna. DSP: digital signal processing. **b**, Image of the transmission link. A mid-way THz reflector doubles the transmission distance within our 60 m-long building. Details of the Tx and Rx (focusing lenses not visible) are shown in the left and right image, respectively.

(c.w.) tone at frequency $f_1 = f_0 + B/2 + \Delta$, where $\Delta = 1 \ldots 2.5$ GHz denotes an additional offset required due to the roll-off of the data signal spectrum. This tone will form the THz local oscillator (LO) for generalized Kramers-Kronig reception. The optical spectrum is shown in Inset 1 of Fig. 3 of the main manuscript. The optical carrier-to-signal power ratio (CSPR) is adjusted by varying the gain of the EDFA while keeping the power of the c.w. tone at optical frequency $f_1$ constant. A second 50/50-coupler is used to add another c.w. tone at optical frequency $f_2 = f_1 - f_{THz}$, which is used as a reference for down-conversion of the data signal and the LO tone to the THz carrier frequency $f_{THz}$. For down-conversion, we use a high-speed uni-travelling-carrier photodiode[1] (UTC-PD), which is designed for operation in the H-band (0.220 … 0.325 THz). The tunable laser sources (Keysight, N7714A) used to provide the optical tones $f_0$ and $f_2$ feature linewidths of less than 100 kHz, and the source (NKT Photonics, Koheras Adjustik) used for generating the LO tone at $f_1$ has a specified linewidth smaller than 0.1 kHz. We adjust the polarization using three polarization controllers (Pol. contr.) to maximize the power after a polarizer (Pol.) to ensure perfect interference of the optical signals. A variable optical attenuator (VOA, Keysight, N7764A) is used to set the optical power level of the overall signal entering the UTC-PD, and the optical spectrum is measured by an optical spectrum analyser (OSA). After down-conversion of the optical signals by the UTC-PD, the THz signals are radiated into free space by a horn antenna (HA) and collimated by a polytetrafluoroethylene (PTFE) lens. After a transmission distance of 110 m, the THz signal is collected by a second lens and a horn antenna. The resulting transmission loss of 19 dB is (over-) compensated by a low-noise amplifier with 24 dB gain, which is coupled to a Schottky-barrier diode (SBD) via a hollow waveguide. The SBD rectifies the THz signal, and a real-time oscilloscope (Keysight UXR0804A) with 80 GHz bandwidth records the SBD output current. Fig. S1b shows photographs of the transmitter, the receiver, and the transmission link. For doubling the distance of 55 m available in our building, a metal plate reflects the THz beam from the transmitter back to the receiver, which is positioned side by side with the transmitter. In our measurement, we did not observe any cross-talk between the THz transmitter and the receiver.

## S2. Model of the Schottky-barrier diode receiver

In this section, we describe the principle and the underlying mathematical model of the SBD envelope detector. The block diagram in Figure S2 illustrates the system model consisting of an idealized THz transmitter (Ideal

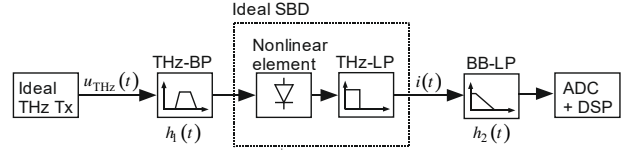

**Fig. S2: Model of the SBD receiver.** An idealized THz transmitter (Ideal THz Tx) generates a perfect THz data signal $u_{THz}(t)$ while a bandpass filter (THz-BP) with impulse response $h_1(t)$ accounts for the bandwidth limitations of the transmitter, of the passive THz components such as antennae and waveguides, as well as of the THz amplifier and the THz circuits at the SBD input. The idealized SBD receiver is described by a nonlinear element followed by an idealized THz low-pass filter (THz-LP). The SBD output current $i(t)$ then enters a baseband low-pass filter (BB-LP) characterized by an impulse response $h_2(t)$, which accounts for the bandwidth limitations of the baseband circuits at the SBD output. The baseband receiver signal is then digitized by an analogue-to-digital converter (ADC), and digital signal processing (DSP) is used for KK phase reconstruction.

THz Tx), a THz band-pass filter (THz-BP) with impulse response $h_1(t)$, a nonlinear element followed by an idealized THz low-pass jointly representing the SBD, an additional low-pass (BB-LP) with impulse response $h_2(t)$ modelling the behaviour of the baseband circuits, and an analogue-to-digital converter (ADC) with subsequent digital signal processing (DSP) for KK phase reconstruction. In this model, the ideal THz Tx is assumed to generate the perfect THz data signal $u_{THz}(t)$ according to Eq. (1) of the main manuscript, while $h_1(t)$ accounts for all impairments due to bandwidth limitations of the various components such as the AWG and the IQ modulator of the optical transmitter, the UTC-PD, the passive THz components such as antennae and waveguides, as well as the THz amplifier and the THz circuits at the SBD input, see Fig. S1. The SBD is modelled by a nonlinear element that represents the relation between the THz voltage and the associated current, followed by an ideal low-pass filter (THz-LP). The role of the ideal THz-LP is to suppress signals at the THz frequency and at harmonics thereof, which are generated in addition to the rectified current $i(t)$ by mathematically applying a nonlinear function to the time-dependent THz voltage $u_{THz}(t)$. In our analysis, the THz-LP is assumed to take the average over several cycles of the THz signal and can be represented by an ideal low-pass filter with cut-off frequency $f_{THz}/2$, see discussion of the mathematical description below. Note that in our consideration, the THz LP is a merely theoretical construct, which is introduced for the sake of a simplified mathematical description. In addition, the baseband circuits at the SBD output are subject to physical bandwidth limitations caused, e.g., by the K-connector used at the SBD package, the subsequent RF cable, and potentially the ADC of the real-time-oscilloscope. The overall effect of these devices is described by another low-pass filter with impulse response $h_2(t)$ and affects the rectified signal $i(t)$.

The effect of the filter with impulse response $h_1(t)$ can be compensated by conventional adaptive feed-forward DSP algorithms applied after KK processing, see Methods for details. The filter with impulse response $h_2(t)$, however, impairs the signal reconstruction according to Eqs. (6) and (14) in the main manuscript[2,3]. The performance of our receiver could hence be further improved by implementing a digital compensation of $h_2(t)$ prior to KK processing[2].

For a mathematical description of the rectification in the SBD, the nonlinear element is described by a function $\tilde{g}$ that connects the THz voltage at the input to a current at the output of the nonlinear element. This connection is assumed to be instantaneous, and the rectified output current $i(t)$ of the ideal SBD is obtained by subsequent averaging over a few THz cycles,

$$i(t) = \langle \tilde{g}(u_{\mathrm{THz}}(t)) \rangle,$$
$$\tilde{g}(u_{\mathrm{THz}}(t)) = \sum_{l=0}^{L} c_l u_{\mathrm{THz}}^l(t). \quad \text{(S1)}$$

In this relation, the time averaging of the ideal THz-LP is denoted as $\langle \cdot \rangle$, and $c_l$ denote the coefficients of a power series expansion of the instantaneous input-output characteristics $\tilde{g}$. According to Eq. (1) of the main manuscript, the THz voltage $u_{\mathrm{THz}}(t)$ incident at the ideal SBD can be written as

$$u_{\mathrm{THz}}(t) = \Re\{\underline{U}(t) e^{j2\pi f_{\mathrm{THz}} t}\}$$
$$= \tfrac{1}{2}\left[\underline{U}(t) e^{j2\pi f_{\mathrm{THz}} t} + \underline{U}^*(t) e^{-j2\pi f_{\mathrm{THz}} t}\right], \quad \text{(S2)}$$

where $\underline{U}^*(t)$ denotes the complex conjugate of the envelope $\underline{U}(t)$. Inserting Eq. (S2) into Eq. (S1) leads to

$$i(t) = \left\langle \sum_{l=0}^{L}\sum_{k=0}^{l} c_l \left(\tfrac{1}{2}\right)^l \binom{l}{k} \underline{U}(t)^{l-k} \underline{U}^*(t)^k e^{j2\pi f_{\mathrm{THz}} t(l-2k)} \right\rangle, \quad \text{(S3)}$$

where $\binom{l}{k}$ denotes the binomial coefficients. Time-averaging suppresses all signal components at the THz carrier frequency and its harmonics $(l-2k)f_{\mathrm{THz}}$ for $l \neq 2k$. Retaining only expressions with $l = 2k$ allows us to express the ideal SBD output current $i(t)$ by the magnitude $|\underline{U}(t)|$ of the complex envelope, equivalently to Eq. (5) of the main manuscript,

$$i(t) = \sum_{k=0}^{K} c_{2k} \left(\tfrac{1}{2}\right)^{2k} \binom{2k}{k} |\underline{U}(t)|^{2k} = g(|\underline{U}(t)|), \quad \text{(S4)}$$

where $K = \lfloor L/2 \rfloor$ with $\lfloor \cdot \rfloor$ denoting the floor function. Note that $g$ represents a bijective function, the inverse of which is expanded into a power series with coefficients $a_n$ according to Eq. (7) of the main manuscript.

### S3. Characterization of SBD receiver

For the generalized Kramers-Kronig processing scheme, we need to reconstruct the THz voltage amplitude $|\underline{U}_R|$ from the measured current $i$. The rectification characteristics of the SBD receiver crucially depend on the operating point as well as on the peripheral THz and baseband circuits and are difficult to describe analytically. We therefore approximate the receiver characteristics $g^{-1}(i)$ according to Eq. (7) of the main manuscript as a power series with $N+1$ initially unknown coefficients $a_n$,

$$|\underline{U}_R| = g^{-1}(i) = \sum_{n=0}^{N} a_n i^n. \quad \text{(S5)}$$

The coefficients $a_n$ are obtained by a least-squares fit of the power series expansion to measured data, which are obtained from the measurement setup shown in Fig. S3a. This setup is a slightly modified version of the setup described in Section S1 without a free-space section between the THz transmitter and the receiver. Note that for this measurement, the unmodulated c.w. tone at $f_1$ is not needed and the associated laser is hence not part of the setup. Note also that the receiver characteristics $g^{-1}(i)$ according to Eq. (S5) do not only account for the nonlinear characteristics of the SBD itself, but also for those of the preceding THz amplifier, see Fig. S3a.

As test signals for characterizing the SBD receiver, we use QPSK signals with a symbol rate of 0.5 GBd. The data pulses have a raised-cosine spectrum with a roll-off factor of 0.1. The symbol rate is intentionally chosen rather small to mitigate the influence of bandwidth limitations of the various devices, which are modelled by band-pass and low-pass filters with impulse responses $h_1(t)$ and $h_2(t)$ in Fig. S2 and which cannot be accounted for by an instantaneous input-output characteristic according to Eq. (S1). Both the transmitted and the received signals are oversampled by a factor of 32, leading to a sequence of known voltages $U_I(t_m)$ and $U_Q(t_m)$ applied to the in-phase (I) and quadrature (Q) arm of the IQ modulator and to a sequence of corresponding SBD currents $i(t_m)$, all measured at discrete times $t_1, t_2, \ldots t_M$. Assuming that the IQ modulator and the UTC-PD are both operated in the linear regime, we can define a complex voltage signal $\underline{U}_{IQ}(t_m) = U_I(t_m) + jU_Q(t_m)$, the magnitude $|\underline{U}_{IQ}(t_m)|$ of which is connected to the magnitude $|\underline{U}(t_m)|$ of the

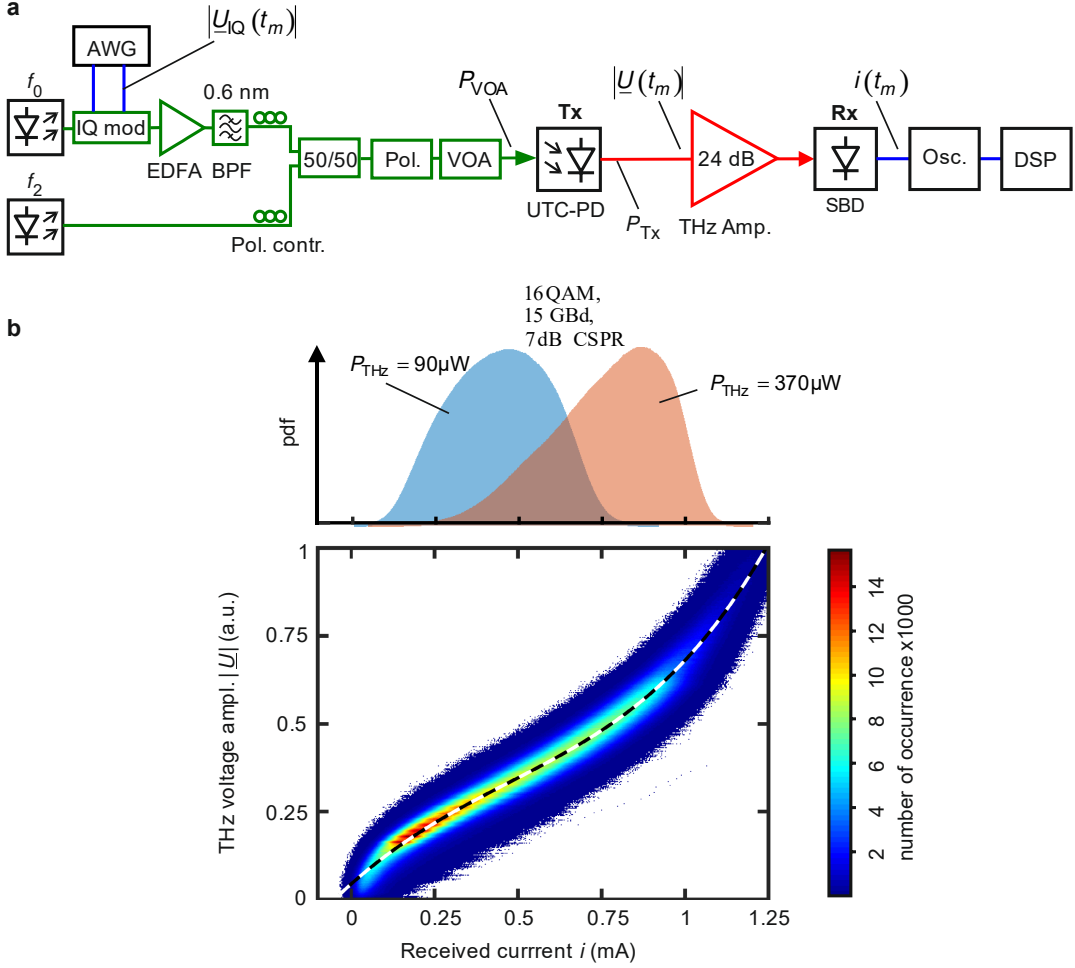

**Fig. S3: Characterization of SBD receiver. a,** Measurement setup used to determine the characteristics of the SBD and the preceding THz amplifier. In the experiment, we feed QPSK test signals with varying THz voltage amplitudes to the input of the THz amplifier and measure the resulting currents at the SBD output. The setup is a slightly modified version of the one described in Fig. S1. Specifically, the free-space section between the THz transmitter and the receiver and the laser generating the unmodulated c.w. tone at optical frequency $f_1$ are omitted. AWG: arbitrary waveform generator. IQ mod: in-phase and quadrature modulator. EDFA: Erbium-doped fibre amplifier. BPF: optical bandpass filter. Pol.: polarizer. VOA: variable optical attenuator. UTC-PD: uni-travelling carrier photodiode. Osc. oscilloscope. DSP: digital signal processing. **b,** Measurement results and fitted SBD characteristics. We record $M = 10^8$ pairs of THz voltage amplitudes $|U| = \xi |U_{MZM}|$ at the input of the SBD receiver along with the associated receiver currents $i$ at the output. The graph in the lower part shows the associated histogram, based on bin sizes of 2.5 µA and 2 mV for the receiver current $i$ and the THz voltage amplitude $|U|$, respectively. The dashed curve shows the least-squares fit of a power series according to Eq. (S5) with $N = 4$. The power series reliably describes the behaviour of the SBD over a broad range of THz voltage amplitudes. The graph on the top shows probability density functions (pdf) obtained from 16QAM data signals transmitted at THz powers of $P_{THz} = 90\,\mu W$ and $P_{THz} = 370\,\mu W$. Even for the low THz power, the vast majority of measured receiver currents are in the region where the fit is reliable.

THz voltage incident at the SBD receiver by a linear relationship,

$$|U(t_m)| = \xi |U_{IQ}(t_m)|. \qquad (S6)$$

Note that, as long as the scaling factor $\xi$ is constant during a measured symbol sequence, its exact magnitude is of no consequence for the Kramers-Kronig receiver, see Section "Phase reconstruction by Kramers-Kronig (KK) processing" in the Methods of the main manuscript.

To cover a broad range of THz amplitudes in our characterization of the SBD receiver, we record time sequences of $|U_{IQ}(t_m)|$ and $i(t_m)$ for different optical power levels at the input of the UTC-PD. These levels are adjusted by a variable optical attenuator (VOA) in front of the UTC-PD, see Fig. S3. Varying the optical output power $P_{VOA}$ of the VOA corresponds to changing the scaling factor $\xi$ in Eq. (S6) in proportion to $P_{VOA}$. We perform a series of measurements at various THz power levels $P_{Tx}$ ranging

from 0.1 µW to 1.4 µW to finally obtain a data set comprising a total of $M = 10^8$ voltage magnitudes $|U(t_m)|$ along with the associated SBD receiver currents $i(t_m)$.

The histogram of this data set is shown in Fig. S3b. The coefficients $a_n$ of Eq. (S5) are obtained by minimizing a cost function $\gamma(a_1, \ldots a_N)$ that is given by the sum of the squared deviations of the measured values $|U(t_m)|$ from the values $|U_R(t_m)|$ reconstructed according to Eq. (7) of the main manuscript,

$$\gamma(a_1, \ldots a_N) = \sum_m^M \left( |U(t_m)| - |U_R(t_m)| \right)^2. \quad (S7)$$

The result for a power series with $N = 4$ is shown as a dashed curve in Fig. S3b. The dashed curve corresponds to the red curve shown in Fig. 2 in the main manuscript. The power series reliably describes the behaviour of the receiver in a broad range of incident THz voltage amplitudes. Only for very small voltage amplitudes, the fitted model function differs from the peak line of the measured histogram. This is a consequence of the fitting procedure, which minimizes the fit error of the THz voltage amplitudes $|U(t_m)|$ and does not account for observational errors of the measured receiver currents $i(t_m)$. A positive effect of this approach is the fact that the resulting function comprises an offset $a_0 \neq 0$, that assigns a zero output current $i = 0$ to a non-zero reconstructed THz amplitude $|U_R| = g^{-1}(0)$. This offset avoids small values of the reconstructed THz amplitude, which would lead to large uncertainties in the phase reconstruction according to Eq. (4) due to the singularity of $\ln(|U_R|)$ at $|U_R| = 0$, and which would otherwise require artificial clipping of the signals at low reconstructed voltages[4]. Note that, in an actual transmission experiment, only a very small share of data points is affected by this offset. This can be understood from typical probability density functions (pdf) of the receiver current, which are exemplarily shown at the top of Fig. S3b. These pdfs refer to 16QAM signals with a symbol rate of 15 GBd transmitted at THz powers of $P_{THz} = 90 \, \mu W$ and $P_{THz} = 370 \, \mu W$, respectively. These THz powers correspond to the leftmost and rightmost point of the plots shown in Fig. 4f of the main manuscript. Even for the smallest THz power, the vast majority of measured receiver currents is in the region where the fit is reliable.

### S4. THz amplifier characterization

We use a low-noise amplifier (LNA) designed for operation in the submillimetre H-band[5] (0.220 THz - 0.325 THz) to compensate the free-space transmission loss and to boost the received THz signal prior to coupling it to the SBD. When characterizing the SBD receiver, Section

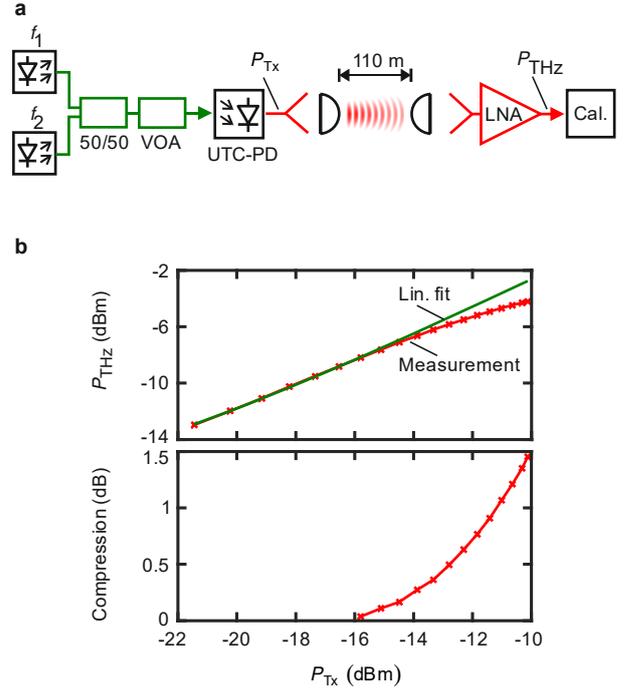

**Fig. S4: THz amplifier characterization. a,** Experimental setup for measuring the THz power after a 110 m-long free-space transmission link and a 24 dB low-noise THz amplifier (LNA). The THz signal at the transmitter is generated by superimposing two unmodulated optical c.w. tones with equal power and by feeding them to a UTC-PD for photomixing. The transmitted THz power $P_{Tx}$ is varied by a VOA which adjusts the optical input power delivered to the UTC-PD. A calorimeter (Cal.) is used to measure the transmitted THz power $P_{Tx}$ as well as the THz power $P_{THz}$ at the LNA output. **b,** Upper graph: THz power $P_{THz}$ at the output of the THz amplifier as a function of $P_{Tx}$. The compression of the amplifier gain leads to a deviation from the linear characteristics. The green curve is obtained from a linear fit of the data points at low power levels $P_{Tx} \leq 16 \, dBm$. Lower graph: Gain compression obtained from the deviation of the measured output power from the power expected according to the extrapolated linear relationship. The maximum compression amounts to 1.5 dB, measured for a transmitted THz power $P_{Tx}$ of -10 dBm.

S3 and Fig. 2 in the main manuscript, we do not discriminate between the nonlinear contribution of the LNA and the SBD because only the combined effect of both is relevant for KK processing. To get a better understanding of the amplifier and its saturation behaviour, we also characterized the device individually. To this end, we use the setup shown in Fig. S4a. Two unmodulated c.w. laser tones having equal powers and different frequencies $f_1$ and $f_2$ are superimposed in a 50/50 coupler and fed to the UTC-PD for photomixing. The resulting THz signal is then radiated to free space, transmitted over 110 m and amplified by the LNA at the receiver. The THz output power after the UTC-PD and after the LNA is measured by a calorimeter (VDI, Erickson PM4).

Fig. S4b shows the measured amplifier output power $P_{THz}$ as a function of the transmitted THz power $P_{Tx}$ measured after the UTC-PD. For large transmitted THz power $P_{Tx}$, the amplifier output power $P_{THz}$ saturates and the measured dependence of $P_{THz}$ on $P_{Tx}$ deviates from the initially linear relationship, see upper graph of Fig. S4b. The corresponding gain compression of the THz amplifier is shown in the lower graph of Fig. S4b. The maximum compression is 1.5 dB. We hence conclude that the compression of the LNA only contributes weakly to the nonlinear receiver response shown in Fig. 2 of the main text.

## S5. Spectral broadening by non-quadratic SBD characteristics

The non-quadratic characteristics of the SBD can lead to a spectral broadening of the rectified output current beyond the bandwidth $B$ of the THz data signal. This effect might increase the bandwidth requirements of the receiver electronics. To investigate this influence, we artificially decrease the receiver bandwidth by introducing a digital $10^{th}$-order Butterworth low-pass filter at the beginning of our digital signal processing chain. As a test, we use QPSK signals with symbol rate $f_{symb} = 30$ GBd and 16QAM signals with symbol rate $f_{symb} = 15$ GBd. The influence of the cut-off frequency $f_{3dB}$ of the low-pass filter on the error-vector magnitude (EVM) of the QPSK signal and on the bit-error ratio (BER) of the 16QAM signal is depicted in Fig. S5a and Fig. S5b, respectively. The insets show the recorded spectrum (blue) of the received signal and the spectrum obtained after digital low-pass filtering (orange) with a cut-off frequency of $f_{3dB} = 1.3 f_{symb}$. Note that the BER and the EVM increase sharply for cut-off frequencies of $f_{3dB} < 1.2 f_{symb}$ because considerable spectral fractions of the down-converted signals are removed for lower cut-off frequencies due to the roll-off of 0.1 that was chosen for raised-cosine pulse shaping.

As expected, the results obtained for heterodyne reception and conventional KK processing are less influenced by low-pass filtering than generalized KK processing, which accounts for the actual non-quadratic characteristics of the SBD and which hence relies on third and higher-order signal-signal mixing products that are partially suppressed by the low-pass. Still, for QPSK signalling, the differences are negligible as long as a filter bandwidths of $f_{3dB} \geq 1.2 f_{symb}$ is used. For 16QAM signals, Fig. S5b, we observe a slight degradation of the signal quality if the cut-off frequency is too small. However, even if the receiver bandwidth is restricted to $f_{3dB} = 1.2 f_{symb}$, the signal quality obtained with the generalized KK receiver is much better than with heterodyne reception or conventional KK processing. We therefore conclude that the effect of spectral broadening is not very prominent and is not a fundamental drawback of generalized Kramers-Kronig processing.

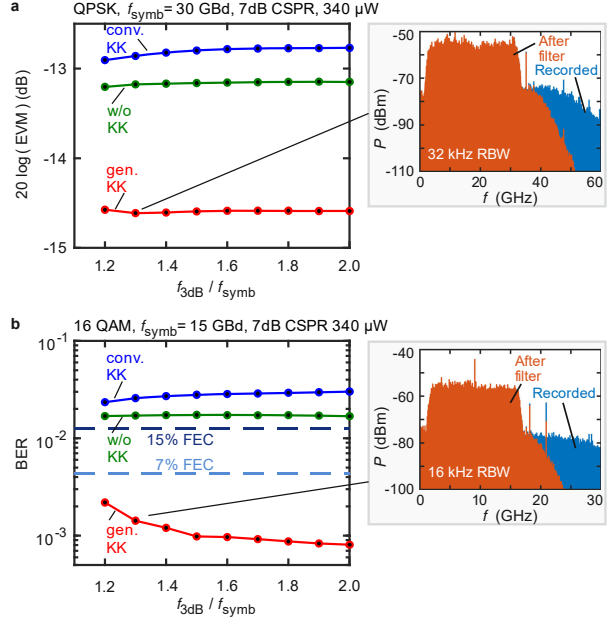

**Fig. S5: Influence of spectral broadening. a,** Error vector magnitude (EVM) as a function of the cut-off frequency $f_{3dB}$ of an artificially introduced low-pass filter for a 30 GBd QPSK data signal. The signal quality is essentially independent of the filter bandwidth as long as $f_{3dB} \geq 1.2 f_{symb}$. The inset shows the recorded spectrum of the receiver current and the spectrum obtained after applying a digital low-pass filter with a cut-off frequency $f_{3dB} = 1.3 f_{symb}$. **b,** Bit-error ratio (BER) as a function of the cut-off frequency $f_{3dB}$ for a 15 GBd 16QAM data signal. The inset shows again the recorded spectrum of the receiver current and the spectrum obtained after applying a digital low-pass filter with a cut-off frequency $f_{3dB} = 1.3 f_{symb}$. For generalized KK processing, a slight increase of the BER can be seen if the receiver bandwidth is not large enough. Still, even for $f_{3dB} = 1.2 f_{symb}$ the signal quality of the generalized KK is much better than with heterodyne reception or conventional KK processing.

## S6. 32QAM measurements

In the main paper, we show the data transmission results for QPSK and 16QAM signals, see Fig. 4. We also investigate the system performance with 32QAM signals. Figure S6 displays the measured BER as a function of the symbol rate and as a function of the incident THz power $P_{THz}$. Also here, the generalized KK scheme shows clear performance advantages over conventional KK processing and heterodyne reception. We reach symbol rates of 5 GBd with a BER below the 7% FEC limit. Note that the resulting net data rate of 23 Gbit/s is significantly lower than the 115 Gbit/s achieved with QPSK and 16QAM modulation. This is partially caused by a non-optimum setting of the oscilloscopes used for sampling the SBD output signal in these experiments. Specifically, the sampling rate was reduced from 256 GSa/s to 128 GSa/s and 64 GSa/s without using a suitable anti-aliasing filter and without increasing the averaging time per sample accordingly. This leads to an aliasing of noise signals and hence to an increase of the noise floor. A more careful choice of device settings could allow for higher symbol rates and better signal qualities. Note that the results shown in Fig. 4b and Fig. 4e of the main manuscript are also slightly impaired by this effect, leading to a slight performance degradation of symbol rates below 20 GBd, for which the sampling rate was reduced to 128 GSa/s without suitable anti-aliasing filter and without increasing the averaging time per sample accordingly, see Methods.

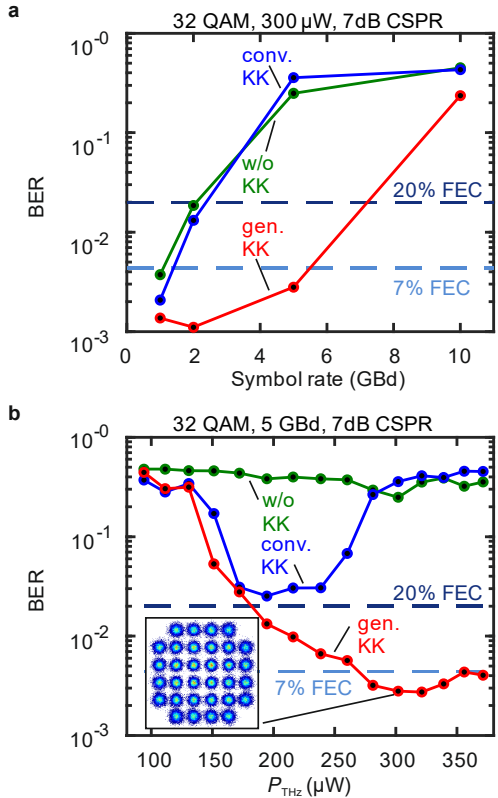

**Fig. S6: Experimental results for 32 QAM measurements. a**, BER as a function of symbol rate with a CSPR = 7 dB and $P_{THz}$ = 340 µW. Generalized KK scheme shows clear performance advantages over conventional KK processing and heterodyne reception. Employing generalized KK reception, the BER stays below the threshold for forward error correction with 7 % FEC overhead for symbol rates of up to 5 GBd. For 5 GBd, we reach a net data rate of 23 Gbit/s. This data rate was limited by a non-ideal setting of the oscilloscopes in our measurement and does not represent a fundamental limit of generalized KK processing. **b**, BER as a function of the incident THz power $P_{THz}$ for a symbol rate of 5 GBd and a CSPR = 7 dB. Also here, the generalized KK scheme shows clear performance advantages over conventional KK processing and heterodyne reception. The constellation diagram for an incident THz power of 300 µW is shown in the inset.